# Analysis of the various key management algorithms and new proposal in the secure multicast communications

Joe Prathap P M.
Department of Information Technology,
Arulmigu Kalasalingam College of Engineering,
Krishnankoil.
Madurai, Tamilnadu, India.
joeprathappm@rediffmail.com

V. Vasudevan
Department of Information Technology,
Arulmigu Kalasalingam College of Engineering,
Krishnankoil.
Madurai, Tamilnadu, India.
drvvmca@yahoo.com

*Abstract*—**With the evolution of the Internet, multicast communications seem particularly well adapted for large scale commercial distribution applications, for example, the pay TV channels and secure videoconferencing. Key management for multicast remains an open topic in secure Communications today. Key management mainly has to do with the distribution and update of keying material during the group life. Several key tree based approach has been proposed by various authors to create and distribute the multicast group key in effective manner. There are different key management algorithms that facilitate efficient distribution and rekeying of the group key. These protocols normally add communication overhead as well as computation overhead at the group key controller and at the group members. This paper explores the various algorithms along with the performances and derives an improved method.**

## I. INTRODUCTION

Multicasting is a type of communication between computers in a network that enables a computer to send one stream of data to many interested receivers without interrupting computers that are not interested. For these reasons, multicasting has become the favored transmission method for most multimedia and triple play applications, which are typically large and use up a lot of bandwidth. Multicasting not only optimizes the performance of your network, but also provides enhanced efficiency by controlling the traffic on your network and reducing the loads on network devices. This technology benefits many group communication applications such as pay-per-view, online teaching, and share quotes [3], [4], [6].

Before these group oriented multicast applications can be successfully deployed, access control mechanisms [2], [9], [13], [19] must be developed such that only authorized members can access the group communication. The only way to ensure controlled access to data is to use a shared group key, known only to the authorized members, to encrypt the multicast data. As group membership might be dynamic, this group key has to be updated and redistributed securely to all authorized members whenever there is a change in the membership in order to provide forward and backward secrecy [5], [8]. Forward secrecy means that a departing member cannot obtain information about future group communication and backward secrecy means that a joining member cannot obtain information about past group communication. We assume the existence of a trusted entity, known as the Group Controller (GC), which is responsible for updating the group key. This allows the group membership to scale to large groups. A number of scalable approaches have been proposed and one in particular, the key tree approach and LKH [2], [3], [10], [19], [20], is analyzed along with its extensions in this paper. In short, the key tree approach employs a hierarchy of keys in which each member is assigned a set of keys based on its location in the key tree. The rekeying cost of the key tree approach increases with the logarithm of the group size for a join or depart request [16], [17], [18]. The operation for updating the group key is known as rekeying and the rekeying cost denotes the number of messages that need to be disseminated to the members in order for them to obtain the new group key.

Individual rekeying, that is, rekeying after each join or depart request, has two drawbacks [12], [14],[19]. First, it is inefficient since each rekey message has to be signed for authentication purposes and a high rate of join/depart requests may result in performance degradation because the signing operation is computationally expensive. Second, if the delay in a rekey message delivery is high or the rate of join/ depart requests is high, a member may need a large amount of memory to temporarily store the rekey and data messages before they are decrypted. Batch rekeying techniques have been recently presented as a solution to overcome this problem. In such methods, a departed user will remain in the group longer and a new user has to wait longer to be accepted. All join and leave requests received within a batch period are processed together at the same time. A short rekey interval does not provide much batch rekeying benefit, whereas a long rekey interval causes a delay to joining members and increases vulnerability from departing members who can still receive the data.

In this paper the existing key management algorithms and its variations were deeply discussed. Based on that a new hybrid key management technique was devised and its performance was analyzed.

## II. BACKGROUND

### A. Simple Algorithm

In the simple group rekeying scheme, all members are connected to a Group Key Controller (GKC). The GKC generates and encrypts the group key separately for every members of the group. When member join/ leave the group, the GKC creates a new key. The GKC encrypts and send this new key to every member separately. Here both communicational and computational complexity is linearly proportional to the group size. This algorithm is not scalable.

### B. Key Tree Approach

In a typical key tree approach [3], [19], [20] as shown in Fig. 1a, there are three different types of keys: Traffic Encryption Key (TEK), Key Encryption Key (KEK), and individual key. The TEK is also known as the group key and is used to encrypt multicast data. To provide a scalable rekeying, the key tree approach makes use of KEKs so that the rekeying cost increases logarithmically with the group size for a join or depart request. An individual key serves the same function as KEK, except that it is shared only by the GC and an individual member.

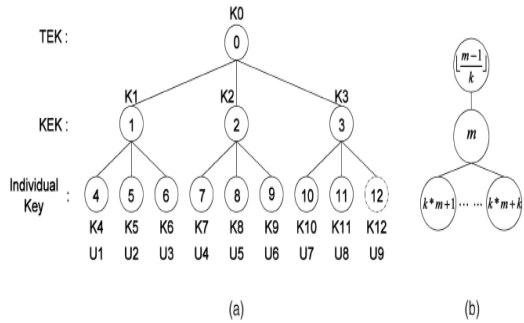

Figure 1. (a) key tree structure (b) ID assignment

In the example in Fig. 1a, K0 is the TEK, K1 to K3 are the KEKs, and K4 to K12 are the individual keys. The keys that a group member needs to store are based on its location in the key tree; in other words, each member needs to store $1 + \log_k N$ keys when the key tree is balanced. For example, in Fig. 1a member U1 knows K0, K1, and K4 and member U7 knows K0, K3, and K10. The GC needs to store all of the keys in the key tree.

To uniquely identify each key, the GC assigns an ID to each node in the key tree. The assignment of the ID is based on a top-down and left-right order. The root has the lowest ID, which is 0. For a node with an ID of m, its parent node has an ID of (m-1)/k, with its children's IDs ranging from km+1 to km+k, as shown in Fig. 1b.

When a member is removed from the group, the GC must change all the keys in the path from this member's leaf node to the root to achieve forward secrecy. All the members that remain in the group must update their keys accordingly. If the key tree is balanced, the rekeying cost for a single departing member is $k\log_k(N)-1$ message. For example, suppose member U9 is departing in Fig. 1a. Then, all the keys that it stores (K0 and K3) must be changed, except for its individual key.

If backward secrecy is required, then a join operation is similar to a depart operation in that the keys that the joining member receives must be different from the keys previously used in the group. The rekeying cost for a single joining member is $2\log_k N$ messages when the key tree is balanced.

The efficiency of the key tree approach critically depends on whether the key tree remains balanced. For a balanced key tree with N leaf nodes, the height from the root to the any leaf node is $\log_k N$. However, if the key tree becomes unbalanced, the distance from the root to a leaf node can become as high as N. and also we can't predict the number of rekeying messages.

### C. Group Key Management Protocol (GKMP)

In GKMP, initially, the GKC selects a member and initiates the creation of a Group Key Packet (GKP). The packet contains the Current Group Traffic Encryption Key (GTEK) and a key (GKEK) to deliver the future GTEK. To handle future rekeys, the GKC then creates a digitally signed Group Rekey Packet (GRP), which consist of the earlier created GKP encrypted with the GKEK. When a member joins, the GKC selects a member and creates a new GKP containing a new GTEK. In addition, it creates a new GRP, which is encrypted under the earlier next GKEK. This method fails to maintain the forward secrecy when a member leaves since every member knows the GKEK[14], [21].

### D. Logical Key Hierarchy

Wong et al. [2] and Wallner et al. [3] independently proposed a scalable group key management scheme by constructing a logical tree of key encryption keys (KEKs) which provides an efficient and secure mechanism to manage the keys and to coordinate the key update. LKH does not take into account the topology of the network. The LKH employs a hierarchical tree whose root node is associated with a group key and whose leaf nodes are individual keys of all users in the group. The intermediate nodes correspond to Key Encryption Key (KEK). Each user in the group holds a set of keys on the path from its leaf to the root[1].

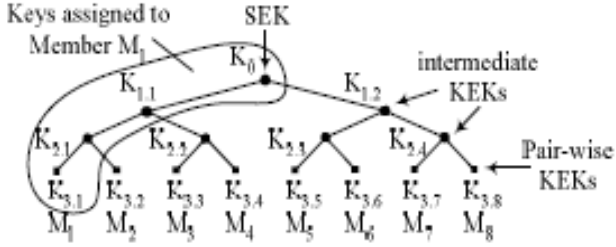

Figure 2.   A binary logical key tree with eight leaf nodes.

Fig. 2 illustrates a rooted binary key tree for a group of eight members. Each member is represented in the tree by a unique leaf node and is preassigned the individual's pair-wise key with the KS. The inner nodes are associated with auxiliary intermediate KEKs, and the root node is associated with the group key SEK. The set of keys associated with the nodes along the path from a leaf node to the root are assigned to the member represented by that leaf node, which include its pair-wise key with the KS, the intermediate KEKs, and the SEK for traffic encryption. For a full and balanced key tree of degree d, the member storage is given as $1 + \log_d N$. For example, member $M_1$ in Fig. 2 is assigned $K_{3.1}$, $K_{2.1}$, $K_{1.1}$ and $K_0$, in which $K_{3.1}$ is $M_1$'s pair-wise key and $K_0$ is the group key.

In the virtual key tree each intermediate KEK can be used to securely multicast rekey messages to members that are leaves of the corresponding inner node's subtree. Member deletion is accomplished by rekeying all the keys possessed by the member (except its pair-wise key) as it shares those with others. For example, when $M_1$ leaves, $K_{2.1}$, $K_{1.1}$ and $K_0$ need to be updated. The number of rekey messages is given as $C_{leave} = d \log_d N − 1$ [7]. In Fig. 2, d =2 and N =8, the KS needs to send the following five rekey messages:

1. $KS \rightarrow M_2$                     :         $\{K'_{2.1}\}K_{3.2}$
2. $KS \rightarrow \{M_2\}$                  :         $\{K'_{1.1}\}K'_{2.1}$
3. $KS \rightarrow \{M, M_4\}$              :         $\{K'_{1.1}\}K_{2.2}$
4. $KS \rightarrow \{M_2, M_3, M_4\}$       :         $\{K'_0\}K'_{1.1}$
5. $KS \rightarrow \{M_5, M_6, M_7, M_8\}$  :         $\{K'_0\}K_{1.2}$

On user joining, each of the involved keys is updated via one unicast (to the new member) plus one multicast (to existing members), requiring a communication cost growing as $C_{join} = 2 \log_d N$. For example in Fig. 2, suppose sometime later $M_1$ joins back the seven-member group, the KS then needs to send the following six rekey messages:

1. $KS \rightarrow M_1$                     :         $\{K'_{2.1}\}K_{3.1}$
2. $KS \rightarrow \{M_2\}$                  :         $\{K'_{2.1}\}K_{2.1}$
3. $KS \rightarrow M_1$                     :         $\{K'_{1.1}\}K_{3.1}$
4. $KS \rightarrow \{M_2, M_3, M_4\}$       :         $\{K'_{1.1}\}K_{1.1}$
5. $KS \rightarrow M_1$                     :         $\{K'_0\}K_{3.1}$
6. $KS \rightarrow \{M_2.....M_8\}$         :         $\{K'_0\}K_0$

In the above approach, each new key is encrypted individually (one key per message) [1].

### E.   One-way Function Tree (OFT)

Canetti *et al.* [4] proposed a variation of LKH by employing a functional relationship among the node keys in a binary key tree along the path from the leaf node representing the leaving member to the root. And this scheme is called as One-way Function Chain(OFC) [9], [18]. OFC reduces the communication overhead from LKH's $2 \log_2 N − 1$ to $\log_2 N$ by introducing a public pseudo-random function G which doubles the size of its input. The left and right halves of G(x) are denoted by L(x) and R(x), so G(x)= L(x)||R(x) where |L(x)| = |R(x)| = |x|. For example, when $M_1$ in Fig. 2 leaves, the KS only sends three rekey messages, from which each residual member can compute all and only the keys it is entitled to receive as depicted in Fig. 3:

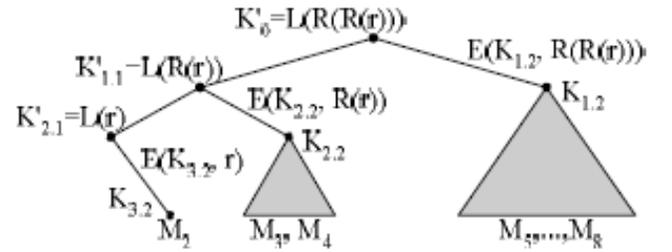

Figure 3.   Key revocation in the one-way function chain scheme

1. $KS \rightarrow M_2$                  :         $\{r\}K_{3.2}$
2. $KS \rightarrow \{M_3, M_4\}$         :         $\{R(r)\}K_{2.2}$
3. $KS \rightarrow \{M_5, M_6, M_7, M_8,\}$  :         $\{R(R(r))\}K_{2.2}$

### F.   Broadcast Encryption-like algorithm

In "broadcast encryption" schemes a central site broadcasts secure transmissions to an arbitrary set of recipients while minimizing key management related transmissions. We take advantage of such methods in order to improve performance for multicast rekeying [8], [16].

The main difference between broadcast and multicast encryption is that the Key Server does not know the identity of possible intruders in broadcast scenarios. In the multicast encryption problem, on the contrary, one knows which of the possible attackers must avoid since the multicast group is limited and known a priori by definition.

### G.   Iterated Hash Chain (IHC)

OFC reduces LKH's communication overhead but it was limited to the binary key tree case. We use LKH's flexible structure to OFC and propose the iterated hash chain (IHC) that supports trees with degree higher than 2 [10], [17]. Fig. 4 presents a ternary key tree, in which there is always a functional relationship among the node keys along the path from the requesting user to the root. A public hash function H is introduced, whose input size and output size both equal the key length in the cryptography system.

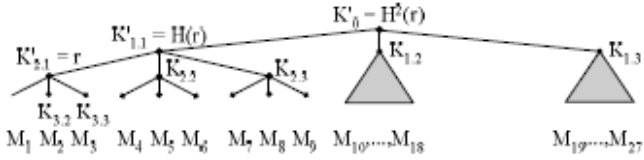

Figure 4.   Key revocation in the iterated hash scheme.

Suppose in Fig. 3 $M1$ leaves the 27-user group. The number of rekey messages is given as $Cleave = (d-1)\log_d N$. The KS chooses a key $r$ at random and rekeys $K2.1$, $K1.1$ and $K0$ as follows:

1.  $KS \rightarrow M2$           : $\{r\}\, K_{3.2}$
2.  $KS \rightarrow M3$           : $\{r\}\, K_{3.3}$
3.  $KS \rightarrow \{M4, M5, M6\}$   : $\{H(r)\} K_{2.2}$
4.  $KS \rightarrow \{M7, M8, M9\}$   : $\{H(r)\} K_{2.3}$
5.  $KS \rightarrow \{M10,....M18\}$   : $\{H^2(r)\} K_{1.2}$
6.  $KS \rightarrow \{M19,....M27\}$   : $\{H^2(r)\} K_{1.3}$

Each residual member can then respectively compute all and only the keys it is entitled to receive as depicted in Fig 4, where the new SEK denoted as $K_0' = H^2(r) = H(H(r))$ is a double iteration. In general we have $K_0' = H^n(r)$ where $n = \log_d N - 1$. Similar to OFC, H should be cryptographically strong so that it is hard to find weak iteration keys.

On user joining, the involved keys are updated via one unicast (to the new member) plus several multicasts (to existing members), leading $Cjoin = 1 + \log_d N$. Suppose sometime later M1 joins back the secure multicast group, the KS then rekeys as follows:

1.  $KS \rightarrow M1$             : $\{r\}\, K_{3.1}$
2.  $KS \rightarrow \{M2, M3\}$      : $\{r\}\, K_{2.1}$
3.  $KS \rightarrow \{M4, M9\}$      : $\{H(r)\}\, K_{1.1}$
4.  $KS \rightarrow \{M10,....M27\}$ : $\{H^2(r)\}\, K_0$

### H.  Synchro-difference LKH (SD-LKH)

It has been generally accepted and recently proven that $O(\log N)$ is the lowest overhead achievable by a key management scheme if strict non-member confidentiality and non-collusion are required [1]. Recent representative improvements and extensions to LKH feature one-way functions [4][5] and clustering [11][12]. In this subsection we present a novel variation that differs observably from those approaches while keeping fully compatible with LKH. The proposal is called synchro-difference LKH (SD-LKH) as new keys are generated based on previous ones by employing the distribution of the difference.

We reuse Fig. 4 to demonstrate the rationale of SDLKH. Suppose $M_1$ leaves the 27-user group. The number of rekey messages is given as $Cleave = (d-1)\log_d N$. The KS randomly chooses a differential value $D$ and transmits the following six messages:

1.  $KS \rightarrow M2$       :      $\{D\}\, K_{3.2}$
2.  $KS \rightarrow M3$       :      $\{D\}\, K_{3.3}$
3.  $KS \rightarrow \{M4, M5, M6\}$      :      $\{D\} K_{2.2}$
4.  $KS \rightarrow \{M7, M8, M9\}$      :      $\{D\} K_{2.3}$
5.  $KS \rightarrow \{M10,....M18\}$      :      $\{D\} K_{1.2}$
6.  $KS \rightarrow \{M19,....M27\}$      :      $\{D\}\, K_{1.3}$

On receiving D, each residual member respectively performs XOR operations to compute all and only the keys it is entitled to receive: $K_{2.1}' = K2.1 \oplus D$, $K_{1.1}' = K_{1.1} \oplus D$, and $K_0' = K_0 \oplus D$. Under the assumption that the generation of $D$ is unpredictable by the users, it should achieve the same security level with LKH.

On user joining, the involved keys are updated via several unicasts (to the new member) plus one multicast (to existing members), leading $Cjoin = \log_d N + 1$. Suppose $M_1$ joins back the group depicted in Fig. 3, the KS then rekeys as follows:

1.  $KS \rightarrow M1$           : $\{K_{2.1}'\} K_{3.1}$
2.  $KS \rightarrow M1$           : $\{K_{1.1}'\}\, K_{3.1}$
3.  $KS \rightarrow M1$           : $\{K0'\} K_{3.1}$
4.  $KS \rightarrow \{M2,....M27\}$     : $\{D\}\, K_0$

in which the new value $(K')$ of each of the keys sent to $M_1(K)$ is computed as $K' = K \oplus D$ by the KS. This still corresponds to one-key-per-message rekeying.

### III.  BATCH REKEYING

Individual rekeying is relatively inefficient, especially when requests are frequent. To address this, the use of periodic batch rekeying was proposed [13] [19]. In batch rekeying algorithms join and leave requests are collected during a time interval and processed in a batch. Since the KS does not rekey immediately, a leaving member will remain in the group till the end of the batch period, and a new member will have to wait the same time to be accepted. However, this batch period can be adapted to dynamics in the multicast group. On the other hand, batch rekeying techniques increase efficiency in number of required messages thus it takes advantage of the possible overlap of new keys for multiple rekey requests, and thus reduces the possibility of generating new keys that will not be used.

### A.  Lam-Gouda batch rekeying

Lam, Gouda et al. [5], [6] presented a very simple marking algorithm that updates the key tree and generates a rekey subtree. Briefly, their system can be summarized as follows. After each rekey interval the KS collects all Join and Leave requests and processes them according to the two possible cases.

If the number of leavings is greater or equal than the number of joinings, new members are allocated in the places of the departed members. Empty leaves are marked as null. All node keys in the path from the replaced leaves to the root are updated following the rules in LKH.

If the number of joinings is greater than the number of leavings a rekey subtree is constructed with all the remaining new members left after applying the algorithm described above. The rekey subtree is allocated under the departed user node with the smallest height.

The algorithm explained in the previous section aims to keep the tree balanced through different batches by allocating the rekey subtree under the shallowest node in each rekeying. However, this rebalancing system is only valid when the number of joinings and leavings are very similar; in any other case a periodic rebalancing algorithm is needed.

### B. Balanced LKH for batch rekeying

In order to overcome this inefficiency we proposed a new batch rekeying algorithm that keeps the tree balanced for every batch**.** The algorithm updates not only node keys but also node naming or position, so rekeying nodes can change their original position after each batch following a very simple rule.

The KS computerized system does not have much more processing load because he only has to update the position of the nodes using simple rules. Besides that, keeping the tree balanced reduces the total amount of required program memory.

In the other side, the new algorithm slightly increases the number of operations to be done by individual members, cause they have to know all the time the position in the tree that they are occupying in order to update it properly. However, this increase is not significant for single multicast members, even if they are devices with low computation capability.

We will briefly describe the atomic steps the KS and the individual members must follow to carry out the algorithm.

#### 1 ) Mark Rekeying Nodes

In the first step, nodes that should be removed have to be pointed out. After collecting the leaving requests, all nodes from leaving members leaves to root need to be updated, **so** they are marked for deletion. Notice that no replacement with joining members is carried out.

#### 2) Prune Tree

The prune action is very simple; it consists in deleting the marked nodes and to keep the subtree structures that remain unchanged. After this action, the KS has to manage three types of elements: remaining subtrees (structures with more than one member), joining members and siblings of leaving members. As the tree is a binary tree, siblings of leaving members cannot reuse any KEK but his individual key, so they should be treated the same way as new joining members.

#### 3) Make New Rekey tree

Now, the **KS** has to construct the new rekey tree balanced following the next recursive criterion. Group all trees of depth j in **twos.** If any element is left, group it with tree of depth j+1 and treat the result as a tree of depth j+2. The criterion must begin with trees of minimum depth, that is to say, single elements, and be repeated until just only one tree is resulted.

#### 4) Construct and Send Rekey Messages

Finally, the rekeying messages have to be sent. These messages should include three information fields: destination node, new position of destination node and rekeying material. The destination node is the node to which sons the message is addressed. This field is used by single members to decide whether the rekeying message concerns to them or not. The new position is the renaming field of the message. Using this information, users can rename themselves and their keying material.

The Rekeying material field is the actual data of updated keys, calculated, for example, according to LKH. On the other hand, the multicast member, basically only has to decide if a multicast rekeying message is sent to him, receive it and update his position and keying material.

### C. Lam-Gouda with Improved LKH

The multicast rekeying consists of three stages: initialization, multicasting and recovering. In the first one the Key Server generates all the secrets to deliver and sends them to the group members. Multicasting stage takes place every time a new shared session key is needed, in this stage the KS sends by the multicast channel the enough data for the authorized members to recompute the new session key, that is to say, the recovering stage [6], [12].

In Lam-Gouda batch rekeying, members are located in the same position of the tree during all the group life. The only changes permitted are up and down the tree level if sibling members leave the group. In any case, the set of keys that each member has from his position to the root is always the same. In such scenario, improved LKH can be straightly applied cause the only information that members need in order to update the keys is $r \oplus r'$. We will better explain the adaptation of Improved LKH to Lam-Gouda Batch rekeying thorough a simple example.

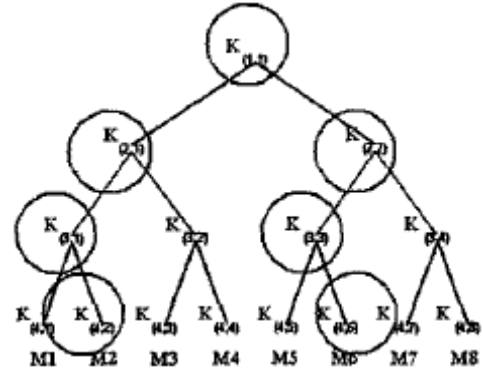

Figure 5. Case of several leaving processed at a time

Consider Fig. *5* in which M2 and M6 leave the group. All the keys that have been compromised (encircled in Fig.5) must be updated. Using Lam-Gouda LKH the length of the next messages should be multicasted is

$$L. \sum_{\log_2 N/L}^{\log_2(N)} l$$

On the other hand, using the improved LKH algorithm the message length is $L.\log_2 N/L$.

### D. Balanced Batch with Improved LKH

In Lam-Gouda [12] with improved LKH we did not allow members to change their position during the group life. That is to say, they always have the same keys in his path to the root.

This allows us the usage of the updating factor $(r \oplus r')$ because members only have to know it to change their set of keys.

Contrary to that, Balanced Batch needs the redistribution of reusable subtrees. This does not permit us only to distribute the updating factor because usually, after a batch, the key path to the root of remaining members will change. This forces us to distribute not the updating factor but the updated keys themselves, although we can use the same mechanism using products of random numbers. This method is expected to present worse behavior than Lam-Gouda with improved LKH. But, on the other side, it keeps the tree balanced all the time. However, performance of such method is better than Balanced LKH without improvement mechanism.

## IV. HYBRID TREE KEY DISTRIBUTION

Since, the number of leaves determines the total number of nodes in a tree of given degree, if we can set the number of leaves as a variable, then we can control the total number of keys. One approach is to cluster the members and assign multiple members to a leaf, then by controlling the number of members assigned to a leaf node, we can vary the total number of nodes in the tree and thus the number of keys stored in the GC. We use the hybrid tree model in to develop the design algorithm for a given amount of update communication.

The main idea of the hybrid tree is to divide the group into clusters of size M with every cluster assigned to a unique leaf node. Then there are N/M clusters (also leaves), and we need to build a tree of depth $\log_a(N/M)$. Fig. 6 illustrates this for a binary tree with cluster size M=3 and a group of 24 members.

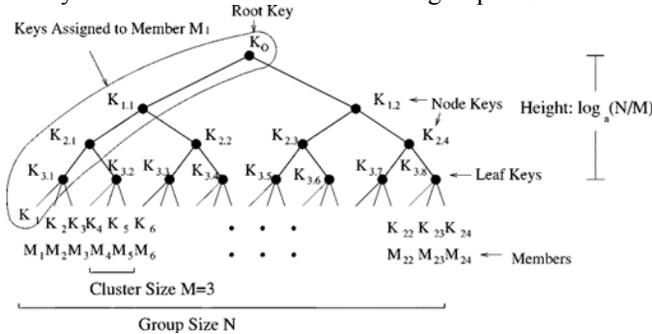

Figure 6. A binary hybrid tree with cluster size M = 3 and group size N = 24

We notice that the structure in Fig. 6 consists of two parts, the logical tree, and the clusters. The logical key tree is used as inter-cluster key management scheme to limit key update communication, and the minimal storage used as the intra-cluster scheme to reduce GC storage requirement.

In the hybrid tree presented in Fig. 6, a user needs to store $(1 + \log_a(N/M))$ KEK's required by the logical key tree scheme plus one KEK required by the minimal storage scheme within the cluster. When a member is deleted, the total number of key update messages, denoted by C , is $(a-1) \log_a(N/M)$ within the tree plus $(M-1)$ within the cluster, leading to:

$$C = (M-1) + (a-1) \log_a(N/M)$$

The number of keys stored by the GC is computed as the keys on the tree plus seeds for ( N/M ) clusters, which is

$$S = \sum_{i=0}^{\log_a(\frac{N}{M})} a^i + \frac{N}{M} = \left(1 + \frac{a}{a-1}\right)\frac{N}{M} - \left(\frac{1}{a-1}\right)$$

The last term $1/(a-1)$ is at most 1 since $a \geq 2$.

Since the logical key tree schemes have logarithmic update communication [1], [2], in the hybrid tree model, we want to keep the update communication as O(log N) except some scale factor $\beta$ . This can be expressed as:

$$(M-1) + (a-1) \log_a \frac{N}{M} \leq \beta \log_a N$$

where the communication scale factor $\beta$ indicates how much communication can be allotted for key updates.

In the hybrid tree scheme, the storage and the update communication are functions of the cluster size M. The selection of M should be such that the update communication scales at least of the order of O(logN) while the key storage of the GC is better than O(N) . Hence the optimization problem is posed as $\min[(1+ a/a-1) N/M]$ w.r.t. M.

With this hybrid tree key distribution the key storage is reduced greater percentage if the total node is in the order of $2^{20}$. The performance is purely based on the cluster size value M. We should choose the cluster value based on the applications & security requirements. Within cluster the communication is very much easier than inter cluster communication. And intra cluster communication provide tight security than the inter cluster communication.

## V. ANALYSIS

The table I provide the comparative analysis of the various algorithms. From this we can easily understand the importance of each algorithm in terms of storage, security, communication costs as well as basic properties. From that we understand that the hierarchical key tree based algorithms work wells for a mildly large group communication networks such as internet radio, video conferences and pay per view systems.

TABLE I.  OVERALL COMPARATIVE ANALYSIS

| Algorithm | Properties | Comm. costs | Secrecy | storage |
|---|---|---|---|---|
| | | | | |

| | Trust in 3rd party | Central controlling entity | Joining & separation of groups | Simple point of failure | Secure against collusions | Scalability | recoverable | Join msg from GKC to other | New member join | Leave from msgs the GKC | Backward | forward | Keys in control manager | Number of member keys | Processing time for retrieving a group key |
|---|---|---|---|---|---|---|---|---|---|---|---|---|---|---|---|
| Simple | X | √ | easy | √ | √ | X | X | n-1 | 1 | n | √ | √ | nk | k | O(logn) |
| GKMP | X | √ | Ok | √ | X | X | X | 2 | 2 | New | √ | X | 2k | 2k | 2k |
| LKH | X | √ | easy | √ | √ | √ | X | d | d+1 | 2d | √ | √ | O(n) | O(logn) | O(logn) |
| OFT | X | √ | easy | √ | √ | √ | X | d+1 | d+1 | d+1 | √ | √ | O(n) | O(logn) | O(logn)/2 |
| Broadcast encryption like algorithm | X | √ | easy | √ | √ | √ | X | d | d+1 | 2d | √ | √ | O(n) | O(logn) | O(logn) |
| IHC | X | √ | ok | √ | √ | √ | X | $1+\log_d n$ | $1+\log_d n$ | $(d-1)\log_d n$ | √ | √ | O(n) | O(logn) | O(logn) |
| SD-LKH | X | √ | easy | √ | √ | √ | X | $\log_d n +1.$ | $\log_d n +1.$ | $(d-1)\log_d n$ | √ | √ | O(n) | O(logn) | O(logn) |
| Lam-Gouda batch rekeying | X | √ | easy | √ | √ | √ | X | d | d+1 | 2d | √ | √ | O(n) | O(logn) | O(logn) |
| Balanced LKH | X | √ | easy | √ | √ | √ | X | d-1 | d+1 | 2d | √ | √ | O(n) | O(logn) | O(logn) |
| Balanced Batch with Improved LKH | X | √ | easy | √ | √ | √ | X | d-1 | d+1 | 2d | √ | √ | O(n) | O(logn) | O(logn) |
| Hybrid Tree Key Distribution | X | √ | easy | √ | √ | √ | √ | $(a-1)\log_a(N/M)$ | N/M +1 | 2N/M | √ | √ | $(a-1)\log_a(N/M)$ | $(1+\log_a(N/M))$ | $(a-1)\log_a(N/M)$ |

## VI.  CONCLUSION

This study exhibits a clear picture about various key management algorithms. The working principles of each algorithm along with their drawbacks are deeply analyzed and tabulated. From this we can easily identify which algorithm is suitable for particular applications. Especially the hybrid key management scheme combine the best features from various algorithms and try to provide an optimized algorithm which is more flexible when compared with other algorithms.

AUTHORS PROFILE

**Joe Prathap P M** received the B.E. and M.E. degree in Computer Science & Engineering from Anna University in 2003 and 2005. He is currently pursuing Ph.D degree in Faculty of Information and Communication Engineering department, Anna University, India. He is currently a senior lecturer at Kalasalingam University. His fields of interest are Computer Networks, Network Security, Operating Systems and Graph Theory. He has more than 10 publications to his credit in international journals and conferences. He is a member of ISTE, India.

**Dr. V Vasudevan** is presently working as Senior Professor and Head, Department of Information Technology at Arulmigu Kalasalingam College of Engineering, KrishnanKoil, Tamilnadu. He is also serving as the Project Director (Network technologies) in TIFAC CORE in Network Engineering at A.K.College of Engineering. He has completed his Ph.D degree from Madurai Kamaraj University, in 199. He has more than 25 years of experience in teaching and research. He has guided more than 50 M.E./M.Tech. projects,10 M.Phil.Thesis and 7 Ph.D. Thesis. His fields of interests are Network Engineering, Multicast Security, Grid Computing and Evolutionary computing. He has more than 50 publications to his credit in international journals and conferences. He has visited many universities in UK. He is a member of ISTE, India.